# Quantum State Engineering and Precision Metrology using State-Insensitive Light Traps

Jun Ye,[1] H. J. Kimble,[2] and Hidetoshi Katori[3]

[1] JILA, National Institute of Standards and Technology and University of Colorado
Boulder, CO 80309-0440, USA
Email: ye@jila.colorado.edu

[2] Norman Bridge Laboratory of Physics 12-33, California Institute of Technology
Pasadena, California 91125, USA
Email: hjkimble@caltech.edu

[3] Department of Applied Physics, School of Engineering, The University of Tokyo
Bunkyo-ku, Tokyo 113-8656, Japan
Email: katori@amo.t.u-tokyo.ac.jp

Precision metrology and quantum measurement often demand matter be prepared in well defined quantum states for both internal and external degrees of freedom. Laser-cooled neutral atoms localized in a deeply confining optical potential satisfy this requirement. With an appropriate choice of wavelength and polarization for the optical trap, two electronic states of an atom can experience the same trapping potential, permitting coherent control of electronic transitions independent of the atomic center-of-mass motion. We review a number of recent experiments that use this approach to investigate precision quantum metrology for optical atomic clocks and coherent control of optical interactions of single atoms and photons within the context of cavity quantum electrodynamics. We also provide a brief survey of promising prospects for future work.

Precision measurement and Quantum Information Science (QIS) require coherent manipulations of electronic states for atoms and molecules with long decoherence times. However, photon recoils create an inevitable back-action on the atomic center-of-mass motion, hence limiting precision and control. In a deeply bound trap, atomic localization within a fraction of an optical wavelength (the Lamb-Dicke regime) greatly reduces motional effects. This capability is exemplified in the Lorentz force-based trapped ion systems with minimal perturbations to internal electronic states. The separation of internal and external dynamics is critical for precision measurement, frequency metrology, and coherent manipulations of quantum systems (*1*).

For neutral atoms, external trapping potentials are created from spatially inhomogeneous energy shifts of the electronic states produced by an applied magnetic, electric, or optical field. In general, such energy shifts are electronic-state dependent, and hence atomic motion leads to dephasing of the two states. A carefully designed optical trap that shifts the energies of the selected states equally provides a solution to this problem.

Light traps employ a.c. Stark shifts $U_i(\vec{r}) = -\frac{1}{2}\alpha_i(\lambda,\varepsilon)|E_L(\vec{r},\lambda,\varepsilon)|^2$ introduced by a spatially inhomogeneous light field $E_L(\vec{r},\lambda,\varepsilon)$, with $\lambda$ the wavelength and $\varepsilon$ the polarization. Two atomic states generally have different polarizabilities $\alpha_i$ ($i$ = 1, 2), resulting in different trapping potentials. A state-insensitive optical trap works at a specific wavelength $\lambda_L$ and polarization $\varepsilon_L$ where $\alpha_1(\lambda_L,\varepsilon_L) = \alpha_2(\lambda_L,\varepsilon_L)$, and $U_1(\vec{r}) = U_2(\vec{r})$ (Fig. 1A). Consequently, the transition frequency $\omega_0$ between the two light-shift-modified

electronic states is nearly decoupled from the inhomogeneous $E_L(\vec{r},\lambda,\varepsilon)$, so long as higher order contributions $O(|E_L|^{n\geq 4})$ are negligible, i.e.,

$$\hbar\omega_0' = \hbar\omega_0 - \frac{1}{2}[\alpha_2(\lambda,\varepsilon) - \alpha_1(\lambda,\varepsilon)]|E_L(\vec{r},\lambda,\varepsilon)|^2 + O(|E_L|^4) \approx \hbar\omega_0.$$

This scenario is possible as $\alpha_i(\lambda,\varepsilon)$ is set by multiple off-resonant atomic transitions. For alkaline earth atoms, the double valence electrons give rise to two distinct series of singlet and triplet states, and the long-lived triplet metastable states are ideal for precision spectroscopy (*2*). In Sr atoms (Fig. 1B), intercombination optical transitions from the ground state $5s^2$ $^1S_0$ to the lowest $^3P_{0,1,2}$ metastable states offer narrow linewidths for clocks. The task then is to find a trapping wavelength for $U_{^1S_0}(\vec{r}) = U_{^3P_{0,1,2}}(\vec{r})$, with negligible scattering losses. For $\lambda > 461$ nm, $\alpha_{^1S_0}$ is always positive, leading to a trapping potential at intensity maximum. For $^3P_0$, the resonances at 2.7 μm and 0.68 μm make the polarizability vary from negative to largely positive as $\lambda$ decreases (Fig. 1C), guaranteeing a match of $\alpha_{^1S_0}$ and $\alpha_{^3P_0}$ at a "magic" wavelength $\lambda_L$ (full curves, Fig. 1D), with its value determined from many relevant electronic states with dipole couplings to $^1S_0$ and $^3P_0$. The shaded curves in Fig. 1D highlight the complexity due to light polarization and the vector nature of an electronic state with angular momentum $J \neq 0$ (e.g., $^3P_1$).

Equalizing light shifts using two different-colored lasers was proposed (*3*) and laser cooling between states of similar polarizabilities in an optical trap was discussed (*4*). To minimize decoherence for quantum-state manipulations, an experimental scheme emerged for a single-wavelength, far-off-resonance dipole trap (FORT) with state insensitivity (*5*). A magic wavelength trap allows (i) two states with the same a.c. Stark shifts, (ii) atoms trapped in the Lamb-Dicke regime, and (iii) atomic center-of-mass motion independent of its internal state (*6*). The experimental realization (*7*) of this proposal (*8*) in strong-coupling cavity QED involving the Cs $6S_{1/2} - 6P_{3/2}$ optical transition led to an extended atomic trap lifetime and the demonstration of diverse phenomena for the interaction of single atoms and photons (*9*). Unlike alkali atoms, intercombination transitions in Sr have linewidths significantly narrower than typical Stark shifts, which critically modify transition dynamics. Efficient cooling on the narrow $^1S_0 - {}^3P_1$ line (*10*) in a state-insensitive optical trap was demonstrated (*11*). An optical lattice clock was proposed using the ultranarrow $^1S_0 - {}^3P_0$ optical transition in $^{87}$Sr (*12*). The use of scalar electronic states ($J = 0$) allows precise control of the Stark shifts solely by the light wavelength, a much better controlled quantity than light intensity or polarization. This is a clear advantage of a state-insensitive trap.

Thus, with independent control of atomic transition and center-of-mass motion, neutral atoms confined in state-insensitive optical traps emulate many parallel traps of single ions, creating greatly enhanced measurement capabilities and new tools for scientific investigations with quantum arrays of atoms and molecules. Two categories of work are progressing rapidly with exciting prospects: (i) precision spectroscopy and frequency metrology (*13-20*), and (ii) quantum-state engineering in the context of cavity QED (*9*).

**Precision Frequency Metrology**

Lasers with state-of-the-art frequency control now maintain phase coherence for 1 s (*21*) and the recent development of optical frequency combs has allowed this optical phase coherence faithfully transferred to other parts of optical or microwave domains (*5*). A new generation of atomic clocks based on optical frequencies, surpassing the performance of the primary Cs standard, has been developed (*20, 22*). A key ingredient is the preservation of the coherence of light-matter interactions enabled by a clean separation between the internal and external degrees of freedom for trapped atoms.

For Sr, the presence of a strong spin-singlet ($^1S_0 - {}^1P_1$) transition and a weak spin-forbidden ($^1S_0 - {}^3P_1$) transition (Fig. 1B and 3A) allows efficient laser cooling in two consecutive stages, reaching high atomic densities and low temperatures limited by photon recoils (<1 µK) (*10, 23*). Transitions between pure scalar states are strictly forbidden. In $^{87}$Sr, nuclear spin $I = 9/2$, and the resulting hyperfine interaction weakly allows the spin- and dipole-forbidden $^1S_0(F = I) \to {}^3P_0(F = I)$ transition ($F$ total angular momentum) with a natural linewidth of ~1 mHz, permitting a high quality factor for the optical resonance (*16*).

*Precision atomic spectroscopy inside a magic-wavelength trap*

With the laser-cooled atoms loaded into a one-dimensional optical standing wave (optical lattice) oriented vertically (Fig. 2), atomic spectroscopy of the $^1S_0 - {}^3P_0$ superposition probes the light-matter coherence at ~1 s. The probe is aligned precisely parallel to the lattice axis to avoid transverse excitations. The Doppler effect is quantized by the periodic atomic motion and is removed via resolved-sideband spectroscopy where the trap frequency far exceeds the narrow transition linewidth. When the probe laser is frequency scanned, a carrier transition appears without change of the motional state. Blue and red sidebands result from corresponding changes of the motional states by ±1 (Fig. 2). The absence of photon recoil and Doppler effects from the carrier transition sets the stage for high precision spectroscopy inside the lattice.

Zooming into the carrier transition, 10 closely spaced resonances are observed with π-excitation (Fig. 3B) under a small bias magnetic field, due to the slightly different Landé $g$-factors between $^1S_0$ and $^3P_0$. This differential $g$-factor, and consequently the hyperfine interaction-induced state mixing in $^3P_0$ and its lifetime, is directly determined from the frequency gap of the resolved transitions (*24*). This high resolution optical spectroscopy measures precisely the nuclear spin effects without using large magnetic fields for traditional nuclear magnetic resonance experiments.

Spin polarization is implemented to consolidate the atomic population to $m_F = +9/2$ or $-9/2$ sublevel. For one particular $m_F$, resonance profiles as narrow as 1.8 Hz (Fig. 3C) are observed, indicating coherent atom–light interactions approaching 1 s. The corresponding resonance quality factor is $2.4 \times 10^{14}$, the highest fractional resolution achieved for a coherent system (*16*). The achieved spectral resolution is limited by the probe laser, with a linewidth below 0.3 Hz at a few seconds and ~2 Hz on 1-minute time scales (*21*).

*Optical atomic clocks*

The concept of a well-engineered trapping potential for accurate cancellation of the differential perturbation to the clock states has led to rapid progress in optical lattice clocks (*13-15*), now demonstrating the high resonance quality factor, high stability (*16, 18*) and low systematic uncertainty (*20*). The high spectral resolution and high signal-to-noise ratio is a powerful combination for precision metrology. Understanding systematic uncertainties of the $^{87}$Sr lattice clock sets the stage for the absolute frequency evaluation by the primary Cs standard via an optical frequency comb. At JILA, this measurement is facilitated by a phase-stabilized fiber link that transfers atomic clock signals between JILA and NIST (*25*), where a Cs fountain clock and hydrogen masers are operating (*26*). Data accumulated over a 24-hour run allow the determination of the $^{87}$Sr $^1S_0 - {}^3P_0$ transition frequency at an uncertainty of $1 \times 10^{-15}$, set by the statistical noise in the frequency comparison (*18*). In Tokyo, the frequency link to Cs reference at NMIJ uses a common view GPS carrier phase technique (*17*). Figure 3D summarizes (*27*) Sr frequency measurements relative to Cs standards in laboratories of Boulder (*14, 18*), Paris (*15, 19*), and Tokyo (*17*). The magic wavelength for the $^{87}$Sr $^1S_0 - {}^3P_0$ transition has been determined independently to be 813.4280(5) nm (*17, 18, 28*) and as expected (*12*), sharing its value at 7 significant digits is sufficient to provide a 15-digit agreement of the clock frequency among the three continents, demonstrating the reproducibility of optical lattice clocks and the success of a new kind of atomic clocks with engineered perturbation.

Under the current operating conditions, the Sr lattice clock has a quantum-projection–noise-limited instability $<1 \times 10^{-15}$ at 1 s, which is somewhat degraded by insufficient stability of the optical local oscillator. With this high measurement precision, rigorous evaluations of the overall uncertainty of an optical atomic clock now demand direct comparison against other stable optical clocks. Stable optical frequencies can be transferred over many kilometers via a phase-stabilized fiber link with stability of $1 \times 10^{-17}/\sqrt{\tau}$ (*25*), permitting evaluation of systematic uncertainties of the JILA Sr clock by remote comparisons against a Ca optical clock at NIST. The overall systematic uncertainty of the Sr lattice clock is currently evaluated near $1 \times 10^{-16}$ (*20*). The low measurement uncertainty achieved in large ensembles of atoms is a powerful testimony to the importance of state-insensitive traps.

**Cavity QED**

An important advance in modern optical physics has been the attainment of strong coupling for the interaction of single atoms and photons. The principal setting for this research has been cavity QED in which an atom interacts with the electromagnetic field of a high-$Q$ resonator in order to investigate fundamental radiative processes associated with the strong interaction of one atom and the electromagnetic field (*5*), with applications in Quantum Optics and Quantum Information Science (*29*).

Various approaches to trap and localize atoms within high-finesse optical cavities have been developed over the past decade with the goal of achieving well-defined coupling $g_0$ between atom and cavity field, where $2g_0$ is the Rabi frequency for a single photon. Beyond atomic confinement *per se*, it is also important that the mechanism for trapping should not interfere with the desired cavity QED interactions for the relevant atomic transitions (e.g., $|b\rangle \leftrightarrow |e\rangle$ in Fig. 4A) (see Section 3 of (*5*)).

The trapping scheme should also support confinement and long coherence times for auxiliary atomic states (e.g., $|a\rangle \leftrightarrow |b\rangle$ in Fig. 4A). For example, the initial proposal for the implementation of quantum networks (*30*) achieves a quantum interface between light and matter via cavity QED. 'Stationary' qubits are stored in the states $|a\rangle, |b\rangle$ and locally manipulated at the nodes of the network. Coherent coupling $g$ to the cavity field and thence to 'flying' qubits between *A*, *B* is provided for one leg of the transition ($|e\rangle \leftrightarrow |b\rangle$), with an external control field $\Omega(t)$ exciting the second leg ($|e\rangle \leftrightarrow |a\rangle$) in a "STIRAP" configuration. Often $|a\rangle$, $|b\rangle$ are hyperfine states (e.g., the "clock" transition $F=3, m_F=0 \leftrightarrow F=4, m_F=0$ in the $6S_{1/2}$ level in Cs), while $|e\rangle$ is an excited electronic state (e.g., in the $6P_{3/2}$ manifold in Cs).

## Cavity QED and the magic wavelength

In contrast to precision metrology where the goal is to isolate a particular atomic transition from external perturbations, strong coupling in cavity QED explicitly introduces large perturbations to the relevant atomic and cavity states. Indeed, for $n$ quanta, the composite eigenstates for a two-state atom coupled to the cavity field experience frequency shifts $\sim \pm \sqrt{n} g(\vec{r})$, as illustrated in Fig. 4B for the $n=1,2$ manifolds. Moreover, in addition to strong coupling for the *internal* degrees of freedom of the atomic dipole and cavity field [i.e., $g(\vec{r}) \gg (\gamma, \kappa)$, with $(\gamma, \kappa)$ the decay rates for atom and cavity], single quanta can also profoundly influence the *external*, center-of-mass degree of freedom, $g(\vec{r}) \gg E_k/\hbar$, with $E_k$ the atomic kinetic energy. Finally, it is possible to interrogate the atom-cavity system at rates exceeding $\gamma \sim 10^8$/s for an allowed dipole transition, with potentially large heating. This situation differs markedly from the more leisurely inquires employed for frequency metrology with a forbidden transition, for which $\gamma \sim 1$/s.

In general, the atom-cavity coupling $g(\vec{r})$ and the ac-Stark shifts $U_e(\vec{r})$, $U_g(\vec{r})$ for excited and ground states (*e*, *g*) have quite different form and magnitude, resulting in a complex spatial structure for the transition frequencies of the atom-cavity system, as discussed in more detail in Section 2 of (*5*). By contrast, in a FORT at $\lambda_L$, $U_e(\vec{r}) \cong U_g(\vec{r}) < 0$, so that the dressed states of the atom-cavity system revert to their basic form $\pm\sqrt{n}g(\vec{r})$ with dependence only on $g(\vec{r})$. From a pragmatic perspective, a great benefit of a FORT operating at $\lambda_L$ is that the powerful techniques for laser cooling and trapping of neutral atoms in free space can be taken over *en masse* to the setting of cavity QED.

*Strong coupling for 1 atom in a state-insensitive trap*

The initial realization of trapping of a single atom inside a high-$Q$ cavity in a regime of strong coupling employed a conventional FORT (i.e., $U_g(\vec{r}) \approx -U_e(\vec{r}) < 0$) with a trap lifetime $\tau \approx 30$ ms (*6*). State-insensitive trapping was achieved later for single Cs atoms stored in a FORT operated at the magic wavelength $\lambda_L$ = 935.6 nm (*7*). The observed lifetime of $\tau \approx 3$ s represented an advance of $10^2 - 10^4$ for trapping in cavity QED (*6, 31*). Moreover, Sisyphus cooling (*32*) for a strongly coupled atom was made possible by $U_e(\vec{r}) \approx U_g(\vec{r})$. Independent investigations of trapping Cs in a free-space FORT around the magic wavelength were reported (*33*).

The combination of strong coupling and trapping at the magic wavelength enabled rapid advances in cQED (*9*). Included are the realization of a one-atom laser in the regime of strong coupling, the efficient generation of single photons "on demand", the continuous observation of strongly coupled and trapped atoms (*7, 34*), and the observation of the vacuum-Rabi splitting $\pm g_0$ (*35*). The experiment in (*35*) (Fig. 4C) is significant in that technical capabilities built around a magic wavelength FORT allowed for a rudimentary quantum protocol with "one-and-the-same" atom., as shown in Fig. 4D. By contrast, all prior experiments related to strong coupling in cavity QED had required averaging over $\sim 10^3 - 10^5$ single-atom trials. Essential components of this work were the state-insensitive FORT and a new Raman scheme for cooling to the ground state of axial motion (*36*). The implementation of complex algorithms in QIS requires this capability for repeated manipulation and measurement of an individual quantum system, as, for example, for the generation of single photons (*37*).

The experimental arrangement depicted in Fig. 4C has also enabled strong photon-photon interactions, as manifest in the phenomenon of photon blockade (*38*). The underlying mechanism is the anharmonicity of the energy spectrum for the atom-cavity system illustrated in Fig. 4B, which arises only for strong coupling and which closely mirrors the free-space structure in a FORT at the magic wavelength. Reversible mapping of a coherent state of light to and from the hyperfine states $|a\rangle, |b\rangle$ of an atom trapped within the mode of a high finesse optical cavity (cf., Fig. 4A) has also been achieved (*39*), thereby demonstrating a fundamental primitive for the realization of cavity QED-based quantum networks (*29, 30*).

*Atomic localization in cavity QED*

Trapping single atoms within high-$Q$ cavities has led to diverse advances in optical physics, including new regimes for optical forces not found in free space (*40-44*). Initially, the principal mechanism for trapping was a red-detuned FORT operated relatively close to atomic resonance, for which $U_e(\vec{r}) \approx -U_b(\vec{r}) > 0$ with correspondingly limited trapping times $\leq 0.1$ s (*6, 43-45*). More recently, $\lambda_F$ has been shifted beyond 1 μm with then $U_e(\vec{r}) < 0$ and much longer trap lifetimes ~10 s achieved (*37, 46*), as well as the deterministic transport of single atoms into and out of the cavity.

Strong coupling with trapped ions is an exciting prospect as the trapping potential for the atomic motion is independent of internal states and trapping times are 'indefinite'. Although great strides have been made (*47, 48*) and the boundary for strong coupling reached (*48*), an inherent conflict is between small mode volume and stable trapping.

## Future prospects

*Precision quantum metrology*

Alkaline earth atoms confined in state-insensitive lattice traps provide a fertile playground for quantum optics and precision measurement-based quantum metrology. Although challenging, the precision of atomic spectroscopy will likely reach the limit set by quantum projection noise. This is an important milestone for large ensembles of atoms and will enable atomic clocks to operate with unprecedented stability. With continued improvement of stable lasers, tomorrow's optical lattice clocks will exhibit instabilities below $10^{-16}$ at 1 s.

Quantum nondemolition measurement for spin-squeezing in an optical lattice can prepare a collective macroscopic pseudo-spin to further enhance the clock stability and precision. High measurement precision will be critical for the evaluation of systematic uncertainties of these new clocks. For example, systematic uncertainties $<1\times10^{-17}$ would require evaluation times of only a few 100 s.

The idea of state-insensitive traps extends to zero nuclear-spin bosonic isotopes of Sr, Yb, or others by using external fields to induce forbidden transitions (*49, 50*). Application of IIb elements (Zn, Cd, and Hg) for optical lattice clocks will significantly reduce the sensitivity to the blackbody radiation-induced shift. Recently, magneto-optical trapping of Hg was reported (*51*). State-insensitive optical traps also benefit research on cold molecules, with important directions towards novel quantum dynamics, precision measurement, and ultracold chemistry. The scalar nature of molecular vibrational levels in the electronic ground state simplifies the search for a magic wavelength for matching polarizabilities between two specific vibrational levels, creating a high-accuracy optical molecular clock (*52*). This molecular system is attractive for searching possible time variations of fundamental constants, particularly the electron-proton mass ratio. Comparison among these different clocks will diversify and strengthen tests of the laws of Nature.

The combination of quantum manipulation and precision metrology in an optical lattice allows accurate assessment of the system's quantum coherence while maintaining precise control of inter-particle interactions. Quantum statistics of nuclear spins can be used to turn on and off electronic interactions. Meanwhile, couplings between nuclear spins in the lattice can be enhanced via electronic dipolar interactions. These electronic interactions are accessed via narrow-linewidth optical Feshbach resonances (*53*) and may allow entangling nuclear spins. These tunable interactions are ideal for QIS where qubits are strongly coupled to one another on demand, but weakly coupled to the error-inducing environment. Furthermore, individual nuclear spins may be addressed and monitored using high spectral resolution optical probes under an inhomogeneous magnetic field. Nonuniform properties of an optical lattice can thus be probed and compensated with spatial addressing.

*Applications of state-insensitive traps in Quantum Information Science*

Recently, quantum degenerate atomic gases have been trapped and strongly coupled to optical cavities (*54-56*) , with a variety of atomic collective effects explored. Another area of considerable activity has been the interaction of light with atomic ensembles (i.e., a large collection of identical atoms), with important achievements reported for both continuous quantum variables and discrete excitations (*57*). In these areas and others, state-insensitive optical traps can enable new scientific capabilities by minimizing the role of decoherence while at the same time allowing coherent optical interactions mediated by electronic excited states. Of particular interest are the implementation of quantum networks and the exploration of the quantum limits to measurement.

**Quantum networks** – Quantum state transfer (Fig. 4A) provides a basis for implementing complex quantum networks (*30*). However, experiments in cavity QED have relied upon Fabry-Perot cavities formed by two spherical mirrors, There have been intense efforts to develop alternative microcavity systems (*58-61*) for scalable quantum networks and quantum information processing on atom chips (*60*). A candidate for trapping individual atoms near a monolithic microcavity is a FORT operated at two magic wavelengths, one red and the other blue detuned from resonance (*62*).

With respect to atomic ensembles (*57*), there is clearly a need to extend coherence times for stored entanglement, where currently $\tau \sim 10^{-5}$ s for entanglement of single excitations between remotely located ensembles. A promising mechanism is confinement of atoms within a state-insensitive trap to realize a long-lived material system for the nodes of a quantum network (*63*). In this setting, dephasing due to position-dependent shifts in transition frequency within the trap is minimized.

**Quantum measurement** – We have previously discussed the prospects for surpassing the limit set by quantum projection noise for precision spectroscopy. In addition to this important possibility, there are other applications of state-insensitive traps to quantum measurement, particularly within the setting of cavity QED. For example, by separating the functions of trapping (via a state-insensitive FORT) and sensing (by way of a probe field in cavity QED), it should be possible to confront the quantum limits for real-time detection of atomic motion, including localization beyond the Standard Quantum Limit. The broader context of such research is that of the dynamics of continuously monitored quantum systems whereby the strong coupling of atom and cavity implies a back reaction of one subsystem on the other as a result of a measurement (*64*).

65. We gratefully acknowledge C. J. Hood, K. Birnbaum, A. Boca, A. D. Boozer, J. Buck, J. McKeever, R. Miller, C. Nägerl, T. Northup, D. Stamper-Kurn, D. Vernooy, and D. Wilson of Caltech, S. Blatt, M. M. Boyd, G. K. Campbell, S. Foreman, C. Greene, J. L. Hall, T. Ido, T. Loftus, A. D. Ludlow, M. Martin, M. Miranda, J. Thomsen, T. Zelevinsky of JILA, J. Bergquist, S. Diddams, T. Fortier, S. Jefferts, C. Oates, T. Parker of the NIST Time and Frequency Division, and M. Takamoto of Tokyo, M. Imae and F.-L. Hong of NMIJ/AIST for their collaborations and discussions. The work at JILA is supported by NIST, NSF, DARPA, and ONR. Work at Caltech is supported by NSF and IARPA. Work at Tokyo is supported by SCOPE and CREST.

**Figure Captions:**

Fig. 1    (A) Atoms inside an optical field experience energy level shifts from the a.c. Stark effect. When the light field is spatially inhomogeneous (a focused beam with Rayleigh range $z_0$ and diameter $w_0$), a light trap is formed. When the polarizabilities of states |1> and |2> are matched by appropriate choices of the light wavelength and polarization, the optical trap becomes state-insensitive.  (B) Level diagram for Sr atoms. The polarizability of the ground state is determined mainly from the strong $^1S_0 - {}^1P_1$ resonance. The metastable triplet states are coupled to the $^3S$, $^3D$, and $5p^2\,{}^3P$ states, with the dominant interactions given by the specific levels shown.  (C) Wavelength dependence of the $^1S_0$ and $^3P_0$ polarizabilities, given in atomic units via scaling by a factor of $1/4\pi\varepsilon_0 a_0^3$. (D) Wavelength-dependent a.c. Stark shifts for the $^1S_0$, $^3P_0$, $^3P_1$ (m = 0), and $^3P_1$ (m = ±1) states, under various light polarizations and intensity $I_0 \sim 10$ kW/cm$^2$.

Fig. 2    $^{87}$Sr lattice clock. Blue laser light ($^1S_0 - {}^1P_1$) is used to cool and trap strontium atoms at the center of the vacuum chamber. Atoms are further cooled with red light ($^1S_0 - {}^3P_1$) in the second stage. Atoms are then loaded into a state-insensitive, vertical 1D optical lattice made of near-infrared light. Top right: Schematic levels for lattice spectroscopy, where the two electronic states are convolved with the quantized motional states. Bottom right: Line shape of a saturated $^1S_0 - {}^3P_0$ electronic transition and the motional sidebands.

Fig. 3    (A) Simplified level diagram for $^{87}$Sr lattice clock. Both cooling transitions are shown, along with the clock transition. (B) The clock transition under a bias magnetic field. Linear π-transitions with (without) spin polarization are displayed in blue (green).  The inset indicates individual nuclear spin states. After spin polarization, the population resides in a single spin state. (C) High-resolution spectroscopy of the clock π-transition for a single $m_F$ state, showing ultranarrow (Q ~2.4 × $10^{14}$) spectrum achieved with a 500 ms Rabi pulse. (D) Recent absolute frequency measurements of the $^{87}$Sr clock transition with respect to Cs standards in laboratories of JILA (circles), Paris (triangles), and Tokyo (squares). The frequency is reported relative to an offset frequency $v_0$ = 429,228,004,229,800 Hz. Error bars indicate +/- one standard deviation in systematic uncertainties.

Fig. 4 (A) Illustration of the protocol of Ref. (*30*) for the distribution of quantum states from system *A* to system *B* by way of atom-photon interactions in cavity QED. As shown in inset (*i*), at *A* the external control field $\Omega_1(t)$ initiates the coherent mapping of the atomic state $|\psi\rangle = c_a|a\rangle + c_b|b\rangle$ to the intracavity field by way of the coupling *g* and thence to a propagating pulse via the cavity output mirror with coupling κ. At the second cavity *B*, the control field $\Omega_2(t)$ implements the reverse transformation as in inset (*ii*), with the incoming pulse from *A* coherently transformed back to $|\psi\rangle$ for the atom at *B*. By expanding to a larger set of cavities connected by fiber optics, complex quantum networks can be realized. (B) Level diagram for the atom-cavity system showing the lowest energy manifolds with *n* = 0, 1, 2 for an atom of transition frequency $\omega_A$ coupled to a cavity with resonance frequency $\omega_C$, with $\omega_A = \omega_C \equiv \omega_0$. Displayed is the eigenvalue structure for the $(6S_{1/2}, F = 4, m_F) \leftrightarrow (6P_{3/2}, F' = 5, m_F')$ transition in Cs (corresponding to $|g\rangle \leftrightarrow |e\rangle$ in (A)) for coupling with rate $g_0$ to two degenerate cavity modes with orthogonal polarizations. The basis for photon blockade for an incident probe field of frequency $\omega_p$ is the suppression of two-photon absorption for the particular detuning $\omega_p$ shown by the arrows. Single photons are transmitted for the transition from ground to lowest excited manifold (i.e., $n = 0$ to $n = 1$), but photon pairs are "blocked" because of the off-resonant character of the second step up the ladder (i.e., $n = 1$ to $n = 2$) (*38*). (C) Experimental arrangement for trapping one atom with an intracavity FORT operated at the magic wavelength $\lambda_L$=936nm for one mode of the cavity and driven by $\varepsilon_{FORT}$ (*32*). Cooling of the radial atomic motion is accomplished with the transverse fields $\Omega_4$, while axial cooling results from Raman transitions driven by the fields $\varepsilon_{FORT}$, $\varepsilon_{Raman}$. The cavity length $l = 42$ $\mu m$ and waist $w_0 = 24$ $\mu m$. Cavity QED interactions take place near a second cavity mode at $\lambda_0$=852nm. (D) Transmission spectrum $T_1(\omega_p)$ and intracavity photon number $\langle n(\omega_p) \rangle$ versus frequency $\omega_p$ of the probe beam $\varepsilon_p$ for an individual strongly coupled atom as in (C) (*35*). $T_1(\omega_p)$ is acquired for 'one-and-the-same atom,' with the two peaks of the 'vacuum-Rabi spectrum' at $\omega_p/2\pi = -20, +32$ MHz in correspondence to the splitting for the lower ($n = 1$) manifold of states in (B). The asymmetry of the spectrum arises from tensor shifts of the $m_F$ excited states in the FORT. The small auxiliary peaks are from the distribution of Clebsch-Gordon coefficients for the $(6S_{1/2}, F = 4, m_F) \leftrightarrow (6P_{3/2}, F' = 5, m_F')$ transitions. The full curve is from the steady-state solution to the master equation (*35*). Error bars represent +/- one standard deviation from the finite number of recorded photo-counts.

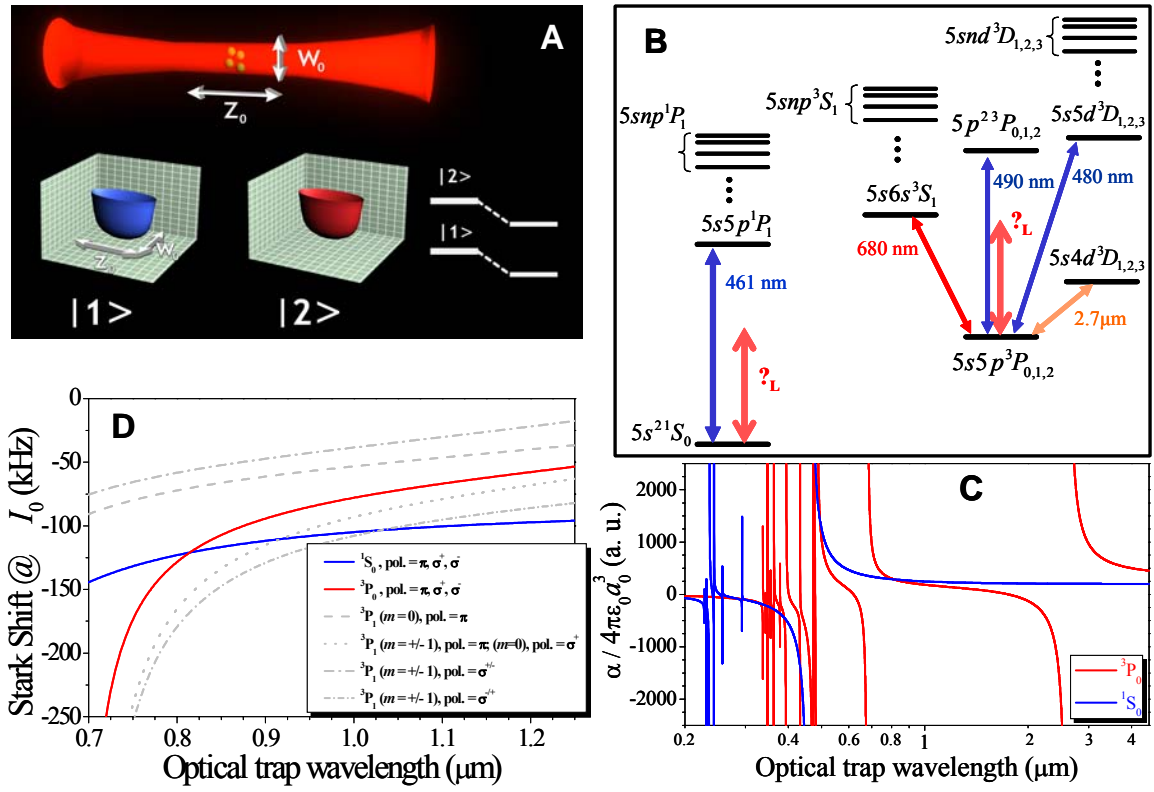

Fig. 1

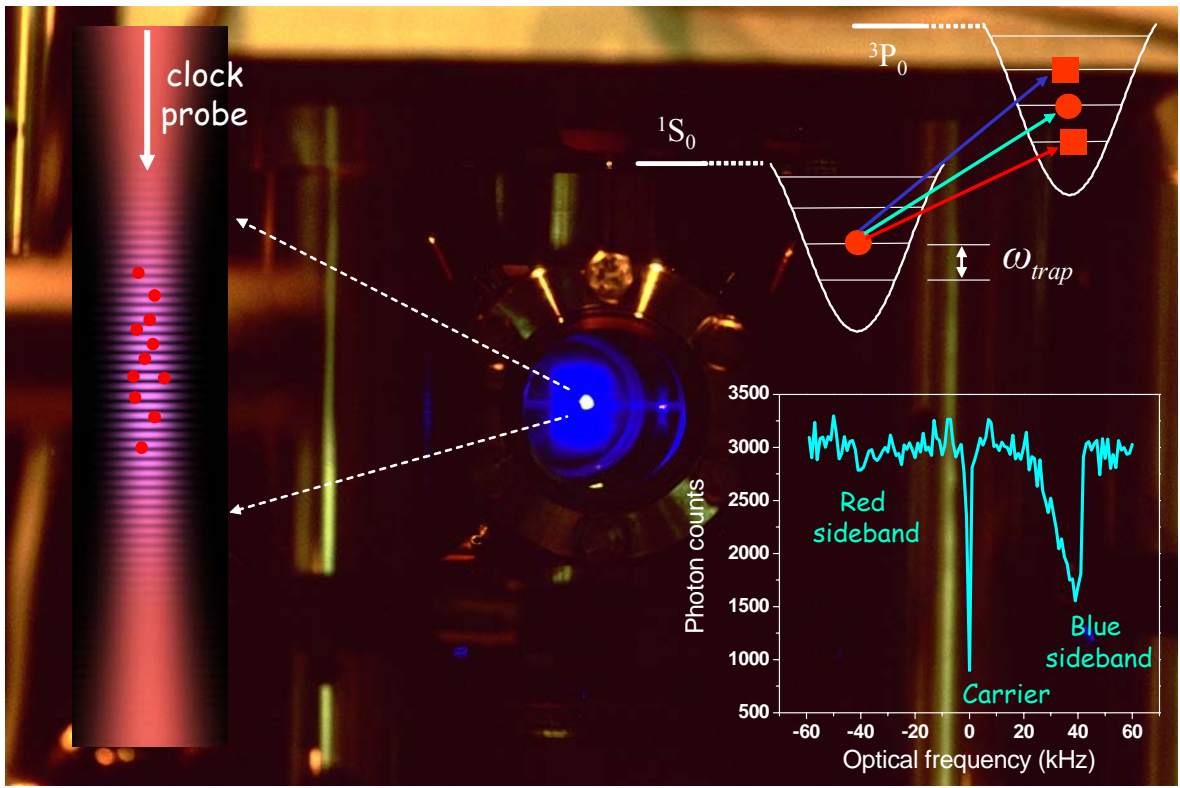

Fig. 2

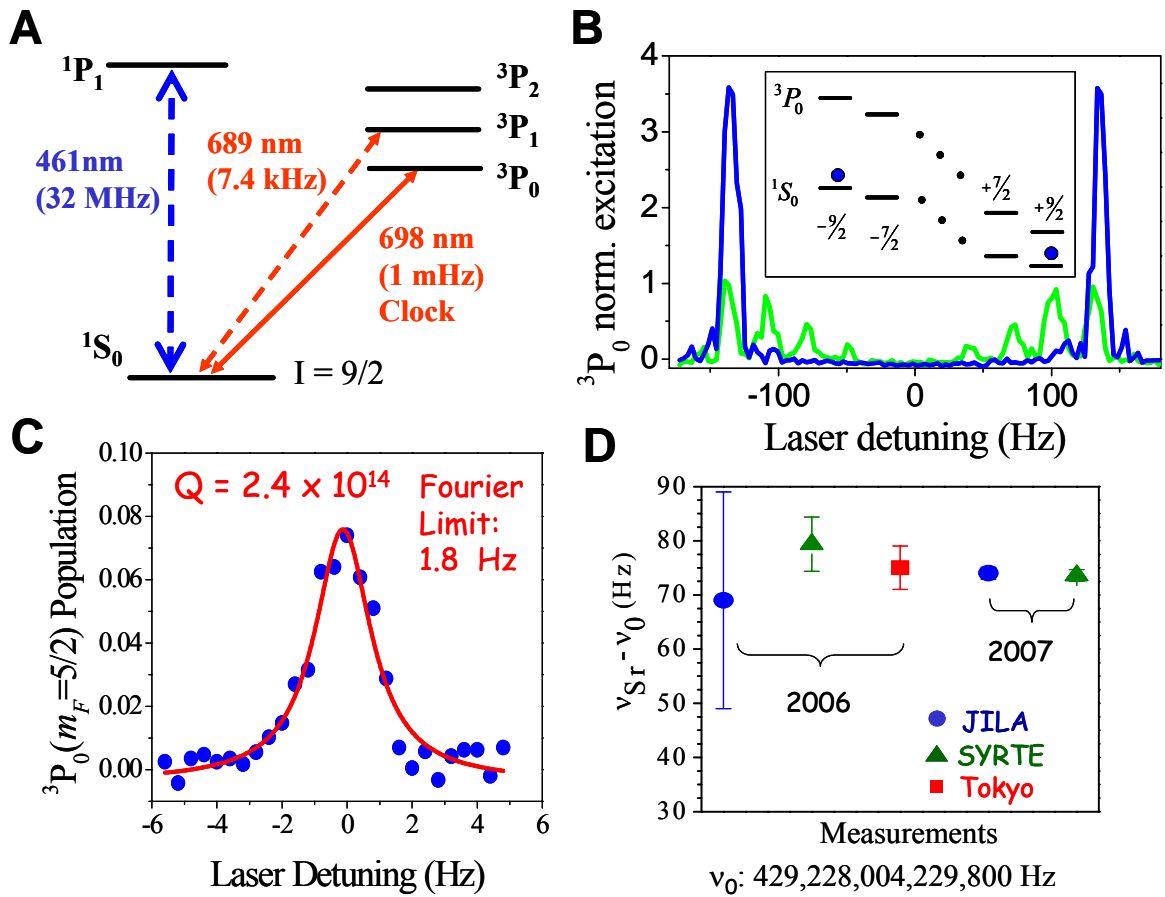

Fig. 3

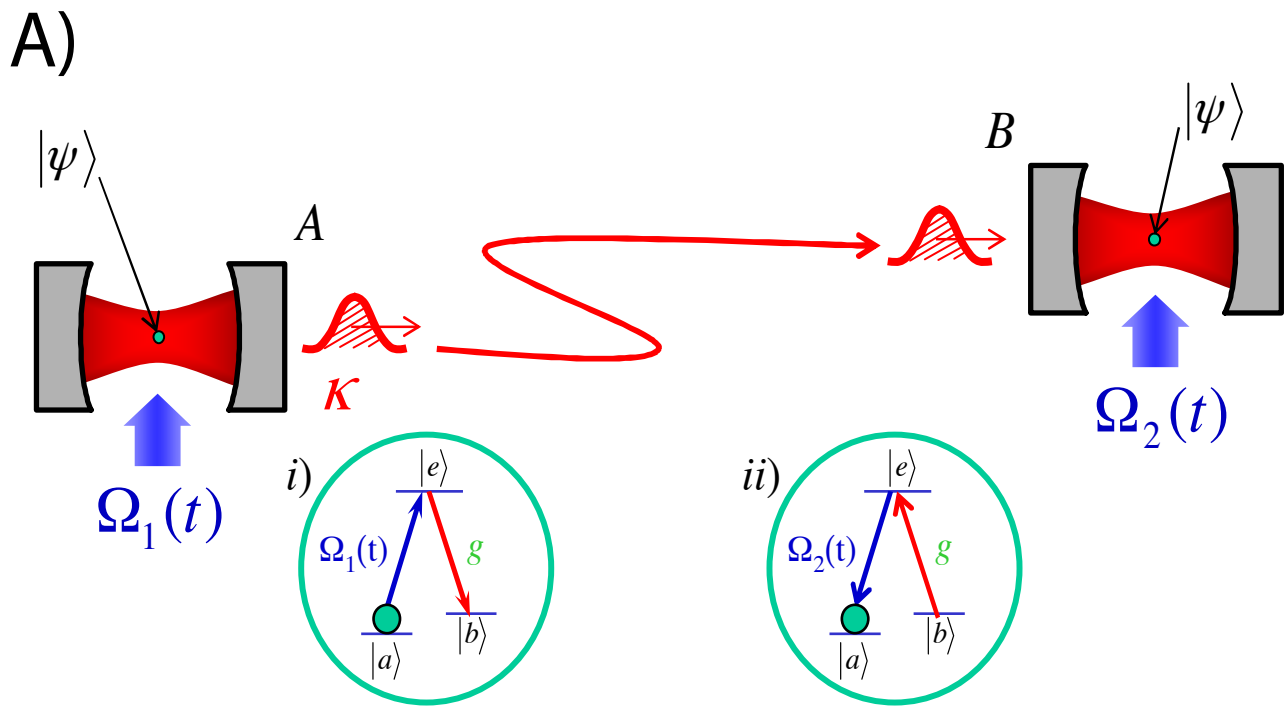
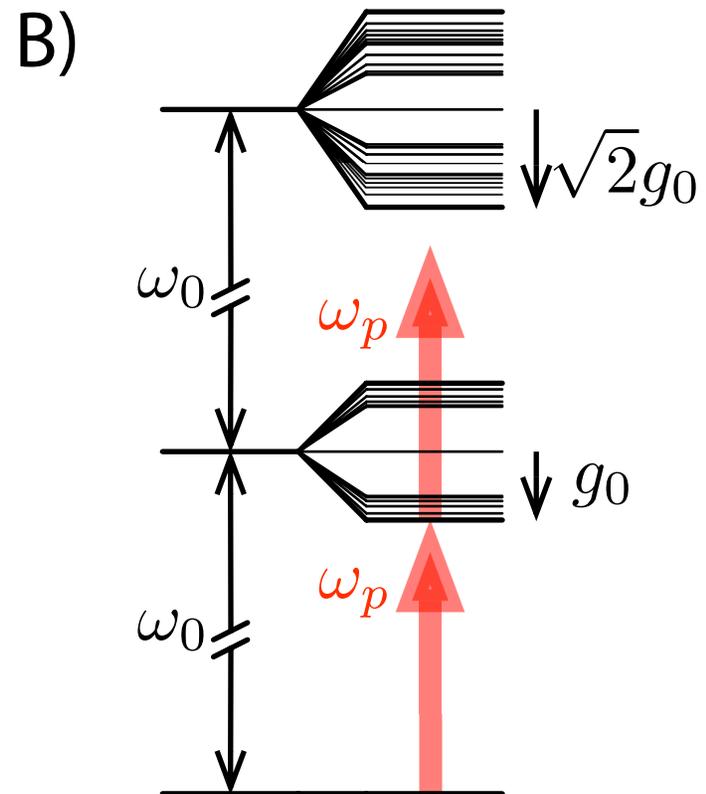
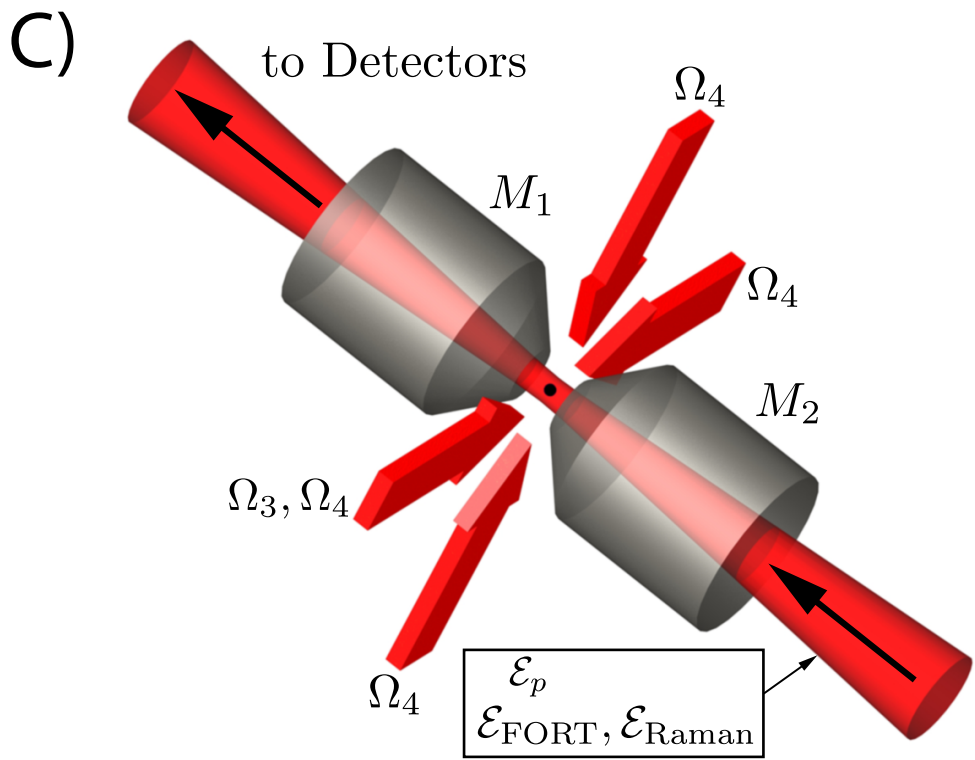
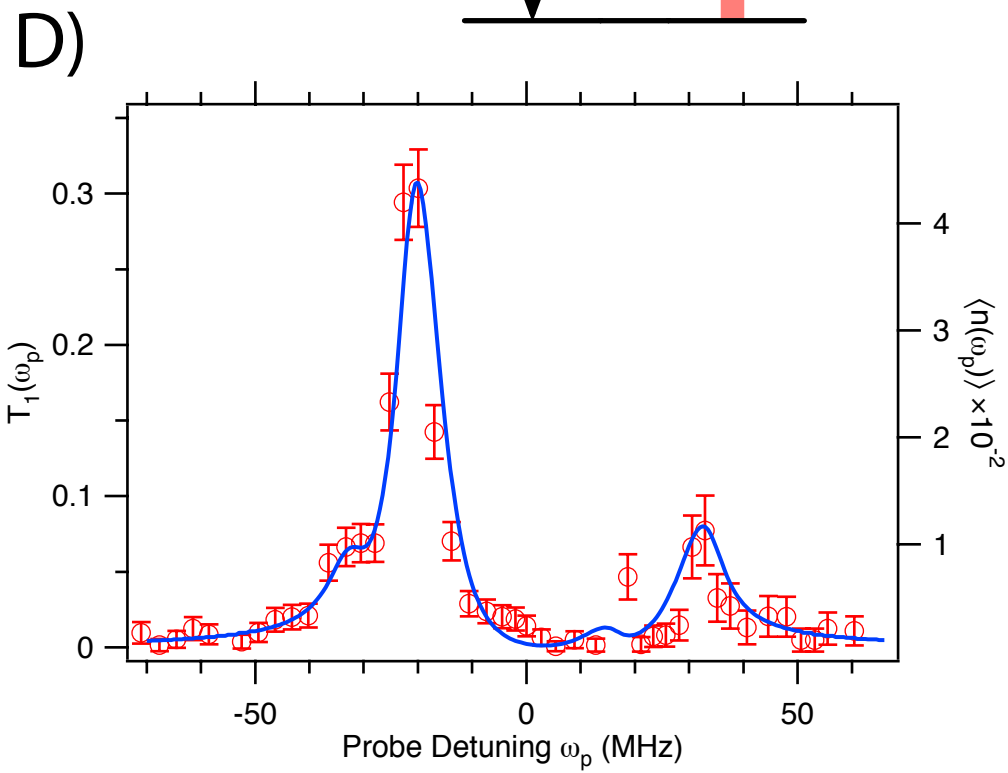

# Supporting Online Material –

# Quantum State Engineering and Precision Metrology

# using State-Insensitive Light Traps

Jun Ye,[1] H. J. Kimble,[2] and Hidetoshi Katori[3]

[1] JILA, National Institute of Standards and Technology and University of Colorado
Boulder, CO 80309-0440, USA
Email: ye@jila.colorado.edu

[2] Norman Bridge Laboratory of Physics 12-33, California Institute of Technology
Pasadena, California 91125, USA
Email: hjkimble@caltech.edu

[3] Department of Applied Physics, School of Engineering, The University of Tokyo
Bunkyo-ku, Tokyo 113-8656, Japan
Email: katori@amo.t.u-tokyo.ac.jp

Supporting documentation is provided for the manuscript Ref. (*1*).

# 1. Preserving the coherence of light – matter interactions

Improvement of spectroscopic resolution has been a driving force behind many scientific and technological breakthroughs, including the invention of laser and the realization of ultracold atoms. The recent development of optical frequency combs has greatly facilitated the distribution of optical phase coherence across a wide range of electromagnetic spectrum. Many excellent references on optical frequency combs have appeared, including (*2-5*). For the state-of-the-art performance in optical phase transfer and comparison, see (*6, 7*).

To preserve the coherence of light-matter interactions, control of the atomic center-of-mass wavefunction is equally important as for the internal states. Trapped ions enjoy the benefit of deep potentials for tight localization of the center-of-mass wavefunction, while the traps normally do not perturb the internal atomic states used for spectroscopy or quantum information processing (*8*). For neutral atoms, the realization of state-insensitive optical traps allows many individual atoms be trapped under a condition like an ion trap. Indeed, experiments reported in (*9*) demonstrate that the level of measurement uncertainties achieved with neutral atom systems can now rival trapped ions. The use of many atoms in neutral systems allows for strong enhancement of the collective signal-to-noise ratio, thereby creating a powerful paradigm to explore precision metrology and quantum measurement and control. Early developments on the magic wavelength optical trap were paralleled in the Caltech group (*10*) and the Tokyo group (*11, 12*). For detailed calculations of magic-wavelength for the Sr optical clock, please refer to (*13*) and (*14*).

Sr atoms are precooled to μK temperatures before they are loaded into an array of optical traps, a one-dimensional optical lattice, formed by an optical standing wave with its axis oriented in the vertical direction. The resulting potential difference between neighboring lattice sites removes the degeneracy of the otherwise translation-symmetric lattice. The formation of localized Wannier-Stark states strongly inhibit tunneling between lattice sites, eliminating a potential problem of accuracy for the optical lattice clock (*15*).

Although both clock states have electronic angular momentum $J$=0, the nuclear spin $I$=9/2 permits ten nuclear spin sublevels, all of which are populated in the ground clock state after cooling. However, a single spin state can be easily achieved by optical pumping. The Stark shifts cannot be completely compensated for all of the magnetic sublevels simultaneously. Or equivalently, the magic wavelength varies slightly for different sublevels. Typically, for the 1D lattice, the laser polarization is linear and coincides with a transverse magnetic field (if it is used to lift the spin degeneracy) to jointly define the quantization axis. Under this configuration, the nuclear spin-dependent vector light shift or the linear Zeeman shift is canceled by averaging the frequencies of a pair of transitions from opposite-signed magnetic sublevels, e.g., $m_F = \pm 9/2$ (*16-18*). The tensor light shift is the same for $m_F = \pm 9/2$ and its effect is thus absorbed into the scalar polarizability that defines the magic wavelength for the ±9/2 spin states.

The typical lattice trap depth is 30-50 photon recoil energy, sufficient to confine atoms in the Lamb-Dicke regime, as the axial trap frequency (tens of kHz) far exceeds the photon recoil (5 kHz), resulting in recoil-free atomic absorptions (*19*). The typical atomic density ranges from $10^{11}$ cm$^{-3}$ to $10^{12}$ cm$^{-3}$. The laser probe is aligned precisely parallel to the lattice axis to avoid transverse excitations and the probe polarization is parallel to that of the lattice laser. The Doppler effect is manifested as modulation sidebands of the unshifted atomic transition (carrier transition) and it is removed completely via resolved-sideband spectroscopy in which the trap frequency is much greater than the narrow linewidth of the clock transition probed by a highly coherent laser. The use of the magic wavelength allows atoms confined in the perturbation-free lattice to preserve the coherence of the $^1S_0$ and $^3P_0$ superposition for 1 s (*20*).

## 2. Level Structure for Cavity QED in a FORT

Altogether, there is a nontrivial set of constraints that should be satisfied for a suitable trapping mechanism in cavity QED, including the possibility for efficient cooling of atomic motion. The important benefits from operation at $\lambda_L$ are clarified from a more detailed examination of the energy level structure for one atom trapped in a cavity in a regime of strong coupling. There is correspondingly a complex interplay of the atom-cavity coupling $g(\vec{r})$ and the ac-Stark shifts $U_g(\vec{r})$, $U_e(\vec{r})$ for ground and excited electronic levels $(g,e)$.

For an atom trapped by a FORT with wavelength $\lambda_F$, denote the ac-Stark shifts for the ground and excited levels $g,e$ by $\delta_{g,e}(\vec{r}) = U_{g,e}(\vec{r})/\hbar$. With reference to Fig. 4(a) in (*1*), assume that the lower manifold $g$ consists of two levels $a,b$ (e.g., hyperfine levels) with equal FORT shifts $\delta_g(\vec{r})$ but with only level $b$ coupled to the cavity mode via the excited state $e$. That is, the atom-cavity coupling $g(\vec{r})$ refers to the $b \leftrightarrow e$ transition as in Fig. 4(a), with the $a \leftrightarrow e$ transition having negligible coupling, which is a good approximation for many experiments.

It is then straightforward to find the position-dependent eigenvalue structure for the atom-cavity system, which consists of a ladder of states with successive rungs $...,n-1,n,n+1,...$, where $n = 0,1,2,...$ gives the number of quanta of excitation shared between atom and cavity field (*21*). The transition frequencies from the ground state with no excitation ($n=0$) to the first excited manifold with two states and 1 quantum of excitation ($n=1$) are given by

$$\Delta^{\pm}(\vec{r}) = \frac{1}{2}(\delta_e(\vec{r}) - \delta_b(\vec{r})) \pm [\frac{1}{4}(\delta_e(\vec{r}) - \delta_b(\vec{r}))^2 + g(\vec{r})^2]^{1/2}, \qquad (1)$$

where $\Delta^{\pm}(\vec{r})$ is measured relative to the "bare," free-space atomic resonance absent the FORT (i.e., the actual optical frequencies are $\omega^{\pm}(\vec{r}) = \omega_A + \Delta^{\pm}(\vec{r})$). Here, we take $\omega_A = \omega_C$ and neglect dissipation $(\gamma,\kappa)$.

For a conventional FORT, $\delta_F^b(\vec{r}) < 0$ thereby providing confinement for an atom in its ground state $b$. However, for the excited state, $\delta_e(\vec{r}) \approx -\delta_b(\vec{r}) \equiv \delta_0(\vec{r})$ leading to (*10, 22-24*)

$$\Delta^{\pm}(\vec{r}) \approx \delta_0(\vec{r}) \pm [\delta_0(\vec{r})^2 + g(\vec{r})^2]^{1/2}. \qquad (2)$$

In general the external trapping potential $\delta_0(\vec{r})$ and the atom-cavity coupling $g(\vec{r})$ have quite different form and magnitude, resulting in complex spatial structure for $\Delta^\pm(\vec{r})$.

An example of the large variation in $\Delta^\pm(\vec{r})$ along the cavity axis is given in Figs. 4, 5 of Ref. (*24*), with excursions in $\Delta^\pm(\vec{r})$ exceeding even the maximum coupling $g_0$. In this case, probe spectra to record the vacuum-Rabi splitting as in Fig. 4(d) of (*1*) would have a quite different form dominated by the spatial variation in $\Delta^\pm(\vec{r})$ and not by the coupling-induced interaction $\pm g(\vec{r})$. Moreover, measurements that require well-defined values for a probe frequency relative to $\Delta^\pm(\vec{r})$ (e.g., photon blockade as in Fig. 4(b)) would become much more problematic.

This said, we should stress that the variation in $\Delta^\pm(\vec{r})$ in a conventional FORT is not without potential benefits. For example, with dissipation $(\gamma, \kappa)$ incorporated into the analysis, new regimes not found for free-space optical forces arise, including mechanisms for heating and cooling of atomic motion within the setting of cavity QED (*24-28*). Here, excitation is provided by driving either the cavity (near $\omega_C$) or atom (near $\omega_A$).

By contrast, in a FORT operated with $\lambda_F$ near a magic wavelength $\lambda_L$, $\delta_e(\vec{r}) \approx \delta_b(\vec{r}) < 0$, with then (*10, 23, 24*)

$$\Delta^\pm(\vec{r}) \approx \pm g(\vec{r}), \qquad (3)$$

so that the transition frequencies to the dressed states depend only on the location $\vec{r}$ of the atom within the cavity mode $\psi(\vec{r})$ (here, for the $n=1$ manifold, but also for arbitrary $n$). A probe beam therefore monitors directly the physics associated with the coherent coupling $g(\vec{r})$ free from the complexity brought by the spatially dependent detuning $\delta_e(\vec{r}) - \delta_b(\vec{r})$ evidenced in Eq. 2. Admittedly, the atom's equilibrium position $\vec{r}_0$ is determined by the structure of the FORT (via $\delta_{a,b}(\vec{r})$), but it is possible to localize the atom such that $g(\vec{r}_0) \approx g_0$ (*29*).

An important practical advantage of operation at a magic wavelength is that powerful techniques for laser cooling and trapping of neutral atoms in free space can be directly applied to the setting of cavity QED (*30*). Until very recently (*31*), strong coupling had been achieved only in Fabry-Perot cavities, which necessarily have limited geometrical access to the mode volume (*32*) and hence restrictions in the ability to illuminate the atom with external control fields. Having the toolbox of free-space cooling techniques available by way of a FORT at the magic wavelength greater expands the options for cooling within the constraints imposed by cavity QED.

### 3. Strong coupling in cavity quantum electrodynamics

Strong coupling in cavity QED requires $g_0 >> (\gamma, \kappa)$, where $2g_0$ is the one-photon Rabi frequency for the oscillatory exchange of one quantum of excitation between atom and cavity field, $\gamma$ is the atomic decay rate to modes other than the cavity mode, and $\kappa$ is the decay rate of the cavity mode itself (*32*). In this circumstance, the number of photons required to saturate an intracavity atom is $n_0 \sim \frac{\gamma^2}{g_0^2} \ll 1$, while the number of atoms required to have an appreciable effect on the intracavity field is

$N_0 \sim \frac{\kappa\gamma}{g_0^2} \ll 1$.

For a dipole-allowed atomic transition, $g_0$ is given by

$$g_0 = \sqrt{\frac{|\vec{\varepsilon}\cdot\vec{\mu}_{ij}|^2 \omega_C}{2\hbar\varepsilon_0 V_m}}, \qquad (1)$$

where $\vec{\mu}_{ij}$ is the transition-dipole moment between atomic states $i, j$ with transition frequency $\omega_A$, and $\omega_C$ is the resonant frequency of the cavity field with polarization vector $\vec{\varepsilon}$ and mode volume $V_m$. If we denote the spatial dependence of the cavity mode by $\psi(\vec{r})$, then the interaction energy $\hbar g(\vec{r})$ likewise becomes spatially dependent, with $g(\vec{r}) = g_0 \psi(\vec{r})$ and $V_m = \int d^3 r\, |\psi(\vec{r})|^2$. A photon of energy $\hbar\omega_C$ in a volume $V_m$ has an associated electric field $E_1 \sim (\hbar\omega_C/V_m)^{1/2}$. Thus for strong coupling, very high-$Q$ cavities ($Q \geq 10^8$) of small volume are required (*32*).

## 4. State-insensitive traps for cold molecules

The state-insensitive optical traps can be applied directly to research on cold molecules, which are expected to play increasingly important roles in studies of novel quantum dynamics, precision measurement, and ultracold collisions and chemical reactions. Cold molecules can be created through photoassociation processes using a weak electronic transition. The narrow transition linewidth requires precise and long-duration atom-light interactions. This condition is fulfilled in a state-insensitive trap (*33*). For example, narrow-line photoassociation near the $^1S_0 - {}^3P_1$ dissociation limit in $^{88}$Sr is an ideal system to test theory - experiment correspondence without the complication of nuclear spins. The wavelength of a state-insensitive lattice trap for the $^1S_0 - {}^3P_1$ transition is ~914 nm (*19*), permitting a recoil- and Doppler-free photoassociation process. The 15 kHz natural width of the molecular line can resolve every vibrational level located near the dissociation limit. The combination of a narrow linewidth least-bound state and its strong coupling to the scattering state should allow efficient tuning of the ground state scattering length with the optical Feshbach resonance technique. The other important feature of this narrow-line photoassociation is relatively large Franck-Condon overlapping factors between vibrational levels of the excited and ground electronic potentials. This favorable overlap leads to efficient productions of ultracold ground-state molecules confined in a lattice field, which can then serve as a basic system for precision test of possible time-dependent drifts of fundamental physical constants. The scalar nature of the molecular vibrational levels in the electronic ground potential permits a straightforward search for a magic lattice wavelength where the polarizabilities of two particular vibrational levels match, thus facilitating accurate measurements of the vibrational energy intervals in the ground potential. This molecular clock system is particularly suitable for measurement on possible variations of the proton-electron mass ratio. The expected constraint reaches 1 x 10$^{-15}$/year (*34*), similar to that provided by atomic frequency metrology. However, tests based on molecular vibration frequencies provide more independence from theory models than atomic tests.